\documentclass[final,3p,times]{elsarticle}

\usepackage[utf8]{inputenc}
\usepackage[sumlimits,intlimits,namelimits]{amsmath} 
\usepackage[english]{babel}
\usepackage{epsfig}
\usepackage{epstopdf}
\usepackage{graphicx}
\usepackage{xcolor}
\usepackage{longtable, multirow,tabularx}
\usepackage{geometry}
\makeatletter
\def\fmslash{\@ifnextchar[{\fmsl@sh}{\fmsl@sh[0mu]}}
\def\fmsl@sh[#1]#2{%
  \mathchoice
    {\@fmsl@sh\displaystyle{#1}{#2}}%
    {\@fmsl@sh\textstyle{#1}{#2}}%
    {\@fmsl@sh\scriptstyle{#1}{#2}}%
    {\@fmsl@sh\scriptscriptstyle{#1}{#2}}}
\def\@fmsl@sh#1#2#3{\m@th\ooalign{$\hfil#1\mkern#2/\hfil$\crcr$#1#3$}}
\makeatother

\newcommand{\hadm}[3]{\left\langle #1 \left| #2 \right| #3 \right\rangle}
\newcommand{\ham}{\mathcal H}

\journal{Physics Letters B}

\begin{document}

\begin{frontmatter}

\title{Light-Quark Decays in Heavy Hadrons}

\author[rvt]{Sven Faller}
\ead{faller@physik.uni-siegen.de}

\author[rvt]{Thomas Mannel}
\ead{mannel@physik.uni-siegen.de}

\address[rvt]{Theoretische Elementarteilchenphysik, Naturwissenschaftlich-Technische Fakult\"at, \\
Universit\"at Siegen, 57068 Siegen, Germany}

\begin{abstract}
We consider weak decays of heavy hadrons (bottom and charmed) where the heavy quark acts as a spectator.
Theses decays are heavily phase-space suppressed but may become experimentally accessible in the near future.
These decays may be interesting as a QCD laboratory to study the behaviour
of the light quarks in the colour-background field of the heavy spectator.   \\
\begin{flushright}
SI-HEP-2015-06 \\[0.2cm]
QFET-2015-07 \\[0.2cm]
\end{flushright}


\end{abstract}

\end{frontmatter}

%
%
%
%


\section{Introduction}
\noindent
Weak decays of heavy hadrons play an important role in shaping our understanding of 
heavy quark physics, see \cite{Manohar:2000dt} and references therein.
Aside from the decays where the heavy quark undergoes a weak transition,
there is also a class of decays in which the heavy quark acts as a spectator
and the light quark decays in a weak transition.
Depending on phase space, this can be either $s \to u$ or in one case also $d \to u$ transitions. 

Due to the very small phase space available in this class of decays, for charmed (strange)
and bottom (strange) mesons only semi-electronic decays are possible.
While the small phase space substantially suppresses these decay modes,
making them difficult to be observed, the small phase space allows for solid 
theoretical predictions, since all form factors need to be known only at the non-recoil point. 

For some of the heavy baryons we can have - aside form the semi-electronic and semi-muonic decays - 
also nonleptonic (pionic) modes.  However, due to the small phase space the pion is quite soft 
in the rest frame of the decaying baryon, which will make the observation of these modes quite difficult. 

Since the $s \to u $ and $d \to u$ transitions have been investigated in all detail in 
ordinary beta decays as well as in 
kaon and hyperon decays, there are no expectations to become sensitive 
to any physics effects beyond the standard model in these heavy-flavour conserving weak processes.  
On the other hand,
theses decays could serve as an interesting cross check of our understanding 
of light quark physics, since the heavy quark in all cases acts as a spectator. Thus the 
physics case for an investigation of such processes is to test the behaviour of light-quark systems 
moving in the (static) colour-background of a heavy quark. 

These decays have not yet attracted a lot of attention. However, the pionic heavy-flavour conserving baryonic  
decay modes have been investigated in \cite{Voloshin:2000et,Li:2014ada} where the relation of these decays 
with the hyperon decays are considered. The same decays have been considered using a model in \cite{Sinha:1999tc}.     


In the next section we first gather all the decays which are possible from the viewpoint of phase space 
and discuss the hadronic matrix elements for a weak transition of the light quarks.
It turns out that the fact that we are basically at zero recoil
(i.e. the velocity of the heavy quark does not change) allows to have on 
the one hand normalization statements for the form factors
derived from the flavour symmetry of the light quarks,
on the other hand the heavy quark spin symmetry allows us
to obtain relations between various decays.
We then first discuss the semi-electronic and semi-muonic  decays
for which we can get quite accurate predictions; 
in a second step we look at the pionic decays,
which cannot be predicted that reliably;
however, we obtain a few benchmark numbers from applying naive factorization.  

\section{Heavy flavour conserving weak decays} 
\noindent
Looking at the spectroscopy of the ground state mesons of bottom and charm
we infer that only semi-electronic decays are allowed,
if we assume that the heavy flavour remains conserved. 
The mass difference between the charged and neutral 
$D$ meson allows for a semi electronic decay through a $d \to u$ transition,
all other decays we consider will be induced by an $s \to d$ transition. 

Strange mesons with a heavy flavour can decay semi-electronically through an $s \to u$ transition; 
for the $B_s$-meson decay, the final state can be a $B$- or a $B^\ast$-meson,
while for the $D_s$-meson the only possible final state is a $D$-meson,
since the $D^\ast$ is too heavy.
In all mesonic cases no hadronic decay is possible since the phase space is too narrow.

\begin{table}[t!]
\caption{\label{BaryHF} List of heavy charm and ground-state baryons \cite{Agashe:2014kda}.
Mass for the $\Sigma_b^0$ baryon taken from Ref.~\cite{Guo:2008ns} and
masses for $\Xi_b^{\prime-}$ and $\Xi_b^{\ast-}$ baryons are taken from the
latests LHCb measurement \cite{Aaij:2014yka}.
In the second column we list the total angular momentum $J$ and parity $P$ of the hadron
and in the third column we give the total spin $s_\ell$ of the light degrees of freedom. }  
\vspace*{4mm}
\begin{tabular*}{\textwidth}{l@{\extracolsep{\fill}}lcccr}\hline\hline 
Baryon              & Mass [MeV] & $J^P$   & $s_\ell$ & Quark Content & $(I,I_3)$  \\ \hline
$\Lambda_c^+$       & 2286.46    & $1/2^+$ & $0$      & $c(ud)_0$ & $ (0,0) $  \\
$\Sigma_c^{++} $    & 2453.98    & $1/2^+$ & $1$      & $c(uu)_1$ & $ (1,1) $\\    
$\Sigma_c^{+} $     & 2452.9     & $1/2^+$ & $1$      & $c(ud)_1$ & $ (1,0) $\\   
$\Sigma_c^{0} $     & 2453.74    & $1/2^+$ & $1$      & $c(dd)_1$ & $ (1,-1) $\\ 
$\Sigma_c^{*++} $   & 2517.9     & $3/2^+$ & $1$      & $c(uu)_1$ & $ (1,1) $\\    
$\Sigma_c^{*+} $    & 2517.5     & $3/2^+$ & $1$      & $c(ud)_1$ & $ (1,0) $ \\   
$\Sigma_c^{*0} $    & 2518.8     & $3/2^+$ & $1$      & $c(dd)_1$ & $ (1,-1) $\\  \hline 
$\Xi_c^{+} $        & 2467.8     & $1/2^+$ & $0$      & $c(su)_0$ & $ (1/2,1/2) $\\ 
$\Xi_c^{0} $        & 2470.88    & $1/2^+$ & $0$      & $c(sd)_0$ & $ (1/2, -1/2) $\\ 
$\Xi_c^{\prime +} $ & 2575.6     & $1/2^+$ & $1$      & $c(su)_1$ & $ (1/2,1/2) $\\ 
$\Xi_c^{\prime 0} $ & 2577.9     & $1/2^+$ & $1$      & $c(sd)_1$ & $ (1/2,-1/2) $\\ 
$\Xi_c^{* +} $      & 2645.9     & $3/2^+$ & $1$      & $c(su)_1$ & $ (1/2,1/2) $\\ 
$\Xi_c^{* 0} $      & 2645.9     & $3/2^+$ & $1$      & $c(sd)_1$ & $ (1/2,-1/2) $\\  \hline 
$\Omega_c^0$        & 2695.2     & $1/2^+$ & $1$      & $c(ss)_1$ & $ (0,0) $ \\
 \hline \hline
$\Lambda_b^0$       & 5619.5     & $1/2^+$ & $0$      & $b(ud)_0$ & $ (0,0) $ \\
$\Sigma_b^0$        & 5810.3     & $1/2^+$ & $1$      & $b(ud)_1$   & $ (1,0 ) $\\
$\Sigma_b^{+} $     & 5811.3     & $1/2^+$ & $1$      & $b(uu)_1$   & $ (1,1) $\\    
$\Sigma_b^{-} $     & 5815.5     & $1/2^+$ & $1$      & $b(dd)_1$   & $ (1,-1) $\\
$\Sigma_b^{*0}$     & 5949.3     & $3/2^+$ & $1$      & $b(ud)_1$   & $ (1,0) $\\
$\Sigma_b^{*+} $    & 5832.1     & $3/2^+$ & $1$      & $b(uu)_1$   & $ (1,1) $\\    
$\Sigma_b^{*-} $    & 5835.1     & $3/2^+$ & $1$      & $b(dd)_1$   & $ (1,-1) $\\   \hline
$\Xi_b^{0} $        & 5793.1     & $1/2^+$ & $0$      & $b(su)_0$   & $ (1/2,1/2) $\\ 
$\Xi_b^{-} $        & 5794.9     & $1/2^+$ & $0$      & $b(sd)_0$   & $ (1/2,-1/2) $\\ 
$\Xi_b^{\prime 0} $ &            & $1/2^+$ & $1$      & $b(su)_1$   & $ (1/2,1/2) $\\
$\Xi_b^{\prime -} $ & 5935.02    & $1/2^+$ & $1$      & $b(sd)_1$   & $ (1/2,-1/2) $\\
$\Xi_b^{*0} $       & 5949.3     & $3/2^+$ & $1$      & $b(su)_1$   & $ (1/2,-1/2) $\\
$\Xi_b^{*-} $       & 5955.33    & $3/2^+$ & $1$      & $b(sd)_1$   & $ (1/2,-1/2) $\\\hline
$\Omega_b^-$        & 6048.8     & $1/2^+$ & $1$      & $b(ss)_1$   & $ (0,0) $\\\hline\hline
\end{tabular*} 
\end{table}

Table~\ref{BaryHF} shows the spectroscopy of heavy flavoured baryonic ground states.
From the point of view of the heavy mass limit, the spin of the heavy quark decouples,
making the baryonic ground states particularly simple \cite{Mannel:1990vg,Falk:1991nq}: 
They consist of a heavy quark, acting as a source of a static colour field,
and a system of light degrees of freedom having either spin $s_\ell = 0$ or $1$.    

Out of these many baryons, only the $\Xi_c$, the $\Omega_c$ states as well as the  $\Xi_b$, 
the $\Omega_b$ states can undergo a heavy flavour conserving weak transition. 
Unlike for the mesons, the phase space of the baryonic weak decays allows for a semi-muonic 
as well as for a hadronic decay with a pion in the final state.

\begin{table}[t!]
\caption{\label{HFCons} List of heavy flavour conserving weak decays as discussed in the text. 
The mass difference is $\Delta m = \sqrt{(M-m)^2 - m_\mu^2}$ for the semi-muonic decays and 
$\Delta m = M-m$ for all the other decays.}
\vspace*{4mm}
\begin{tabular*}{\textwidth}{l@{\extracolsep{\fill}}rccr}\hline\hline
  Decay                      & $\Delta m$ [MeV]%
                                                &$J^P\to J^{\prime P^\prime}$ & $s_\ell\to s_\ell'$ %
                                                & Quark Transition  \\\hline 
  \multicolumn{5}{c}{Semi-electronic decays } \\ \hline
  $D^+   \to D^0 e^+ \nu$    & $4.8$            %
                                                &$0^-\to0^-$                   & $1/2\to1/2$%
                                                & $d\to u$  \\
  $D_s^+ \to D^0 e^+ \nu$    & $103.5$          %
                                                & $0^-\to0^-$                  & $1/2\to1/2$%
                                                & $s\to u$  \\
  $B_s^0 \to B^- e^+\nu$     & $87.5$           %
                                                & $0^-\to0^-$                  & $1/2\to1/2$%
                                                & $s\to u$  \\
  $B_s^0 \to B^{*-} e^+\nu$  & $41.6$           %
                                                & $0^-\to1^-$                  & $1/2\to1/2$%
                                                & $s\to u$  \\\hline 
  $\Xi_c^0  \to \Lambda_c^{+}  e^- \bar\nu$
                             & $184.4$          %
                                                & $1/2^+\to1/2^+$              & $0\to0$%
                                                & $s\to u$  \\
  $\Xi_c^0  \to \Sigma_c^{+}  e^- \bar\nu$ 
                             & $18.0$           %
                                                & $1/2^+\to1/2^+$              & $0\to0$%
                                                & $s\to u$  \\
  $\Xi_c^+\to \Sigma_c^{\ast++} e^- \bar\nu$        
                             & $13.8$           %
                                                & $1/2^+\to3/2^+$              & $0\to1$%
                                                & $s\to u$  \\
  $\Omega_c^0  \to \Xi_c^{+}  e^- \bar{\nu}$
                             & $227.4$          %
                                                & $1/2^+\to1/2^+$              & $1\to0$%
                                                & $s\to u$  \\
  $\Omega_c^0  \to \Xi_c^{\prime +}  e^- \bar{\nu}$%
                             & $119.7$          %
                                                & $1/2^+\to1/2^+$              & $1\to1$%
                                                & $s\to u$  \\
  $\Omega_c^0  \to \Xi_c^{* +}  e^- \bar{\nu}$  %
                             & $49.3$           %
                                                & $1/2^+\to3/2^+$              & $1\to1$%
                                                & $s\to u$  \\\hline                       
  $\Xi_b^-  \to \Lambda_b^{0}  e^- \bar{\nu} $
                             & $175.4$          %
                                                & $1/2^+\to1/2^+$              & $0\to0$%
                                                & $s\to u$  \\ 
  $\Omega_b^-  \to \Xi_b^0  e^- \bar{\nu} $%
                             & $255.7$          %
                                                & $1/2^+\to1/2^+$              & $1\to0$%
                                                & $s\to u$  \\
  $\Omega_b^-  \to \Xi_b^{\prime 0}  e^- \bar{\nu} $%
                             &                  %
                                                & $1/2^+\to1/2^+$              & $1\to1$%
                                                & $s\to u$  \\
                            $\Omega_b^-  \to \Xi_b^{* 0}   e^- \bar{\nu} $%
                             & $99.5$           %
                                                & $1/2^+\to3/2^+$              & $1\to1$%
                                                & $s\to u$  \\\hline
  \multicolumn{5}{c}{Semi-muonic decays } \\ \hline  
    $\Xi_c^0  \to \Lambda_c^{+}  \mu^- \bar\nu$
                             & $151.2$          %
                                                & $1/2^+\to1/2^+$              & $0\to0$%
                                                & $s\to u$  \\    
    $\Omega_c^0  \to \Xi_c^{+}  \mu^- \bar{\nu}$
                             & $201.4$          %
                                                & $1/2^+\to1/2^+$              & $1\to0$%
                                                & $s\to u$  \\    
    $\Omega_c^0  \to \Xi_c^{\prime +}  e^- \bar{\nu}$%
                             & $56.1$          %
                                                & $1/2^+\to1/2^+$              & $1\to1$%
                                                & $s\to u$  \\      \hline
     $\Xi_b^-  \to \Lambda_b^{0}  \mu^- \bar{\nu} $
                             & $140.0$          %
                                                & $1/2^+\to1/2^+$              & $0\to0$%
                                                & $s\to u$  \\      
      $\Omega_b^-  \to \Xi_b^0  \mu^- \bar{\nu} $%
                             & $232.8$          %
                                                & $1/2^+\to1/2^+$              & $1\to0$%
                                                & $s\to u$  \\  \hline                                                                                                             
  \multicolumn{5}{c}{ Pionic decays}\\ \hline
  $\Xi_c^0  \to \Lambda_c^{+}  \pi^-$
                             &$184.4$           %
                                                &$1/2^+\to1/2^+$& $0\to0$%
                                                & $s\to u$   \\
  $\Xi_c^+  \to \Lambda_c^{+}  \pi^0$
                             &$181.3$           %
                                                &$1/2^+\to1/2^+$& $0\to0$%
                                                &  $s\to u$  \\
  $\Omega_c^0  \to \Xi_c^{+}  \pi^-$ 
                             & $227.4$          %
                                                &$1/2^+\to1/2^+$& $1\to0$%
                                                & $s\to u$   \\
  $\Omega_c^0  \to \Xi_c^{0}  \pi^0$ 
                             & $224.3$          %
                                                &$1/2^+\to1/2^+$& $1\to0$%
                                                & $s\to u$   \\\hline 
  $\Xi_b^-  \to \Lambda_b^{0}  \pi^-$
                             & $175.4$          %
                                                &$1/2^+\to1/2^+$& $0\to0$%
                                                & $s\to u$   \\   
  $\Xi_b^0  \to \Lambda_b^{0}  \pi^0$
                             & $173.6$          %
                                                &$1/2^+\to1/2^+$& $0\to0$%
                                                & $s\to u$   \\   
  $\Omega_b^-  \to \Xi_b^0  \pi^-$
                             & $255.7$          %
                                                &$1/2^+\to1/2^+$& $1\to0$%
                                                & $s\to u$   \\        
  $\Omega_b^-  \to \Xi_b^-  \pi^0$
                             & $253.9$          %
                                                &$1/2^+\to1/2^+$& $1\to0$%
                                                & $s\to u$   \\\hline\hline                                           
 \end{tabular*}
 \end{table}

Table~\ref{HFCons} lists all possible heavy flavour weak decays for bottom and charm hadrons.
The second column in the table lists the mass differences of the initial
and final state heavy hadrons.  
We note that all mass differences are large compared to the electron mass,
so we can neglect the electron mass in the following, while we have to keep the pion and the muon mass. 
%
%
%
\subsection{Form factors for light-quark currents} \label{ssec:FF} 
\noindent
To describe the decays shown in Table~\ref{HFCons}
we need matrix elements of light-quark currents with heavy hadron states.
The heavy quark is in these decays only a spectator and acts in the 
infinite-mass limit as a static source of colour.
In other words, we need to look at the transition in light-quark system
in the colour background created by the (static) heavy quark. This picture allows us 
to obtain information on the form factors.  

The four-momenta of the initial $H_i$ and final  $H_f$ heavy hadrons
are $p^\mu = Mv^\mu$ and $p^{\prime\mu} = mv^{\prime\mu}$, respectively,
and $q^2 = (p-p')^2$ is the momentum transfer squared from the hadronic to the leptonic systems.
Instead of the momentum transfer squared we use the variable $w = v\cdot v'$,
\begin{equation}
  w = \frac{M^2+m^2-q^2}{2Mm}\ ,
\end{equation}
where the kinematic boundaries are given by
\begin{eqnarray}\label{resw}
  1 \leq w \leq w_{\rm max} &=& \frac{M^2+m^2}{2Mm}
                  =   1+\frac{(M-m)^2}{2Mm} \sim 1 \ ,
\end{eqnarray}
showing that the range of $w$ is tiny for all decays listed in Table~\ref{HFCons},
since in all cases $(M-m) \ll M$. 
Assuming that the form factors are slowly varying functions of the kinematic variables, 
we may replace all form factors by their values at $w=1$. 
Thus in the following we only need to obtain 
some insight into the form factor in the region $v \sim v'$. 

For the mesonic decays we define the relevant form factors as ($q,q' = u,d,s$) 
\begin{eqnarray}\label{mesonFF}
   \frac{\langle H_f (p') | \bar{q}^\prime  \gamma_\mu  q | H_i (p) \rangle}{\sqrt{Mm}} &=&%
      (v  + v')_\mu \Phi_+ (w) + \dots  \ ,  \\ 
   \frac{\langle H_f^\ast (p',\epsilon) | \bar{q}^\prime  \gamma_\mu  \gamma_5 q | H_i (p) \rangle}{\sqrt{Mm}} &=&  
       i(w+1)  \epsilon_\mu^\ast \Phi_{A_1} (w) + \dots  \ ,   \label{mesonFFs}
\end{eqnarray} 
where we only show the form factors relevant for the leading contribution in the limit $v \to v'$.
In Eq.~\eqref{mesonFFs}, $\epsilon_\mu$ is the polarization vector
of the excited final state meson $H^\ast(p',\epsilon)$.
Taking the heavy quark as static, we need to look at the transition of a light state
with the quantum number of the light quark in the meson $H_i$ into the corresponding final light state 
in $H_f$ via the vector and axial-vector (light quark) current.

Furthermore, despite of the heavy quark's colour field,
the light quark system has an $SU(3)_L \times SU(3)_R$ chiral symmetry,
which is generated by the currents in \eqref{mesonFF} and \eqref{mesonFFs}.
However, this symmetry is spontaneously broken to the usual $SU(3)_{L+R}$ flavour 
symmetry of the light quarks. Assuming that this symmetry is exact, we   
derive from the conservation of the vector current the normalization statement
\begin{equation} 
 \Phi_+ (1)  = 1   \ ,
\end{equation}  
while the light-quark flavour symmetry does not tell us anything about $\Phi_A (1)$.  

The case of the baryonic decays is more interesting,
since the light-quark systems are composed of two valence quarks.
For the case of a transition between two ``$\Lambda$-like'' heavy baryons
(i.e.\  baryons in a $h(qq')_0$ configuration) the light quark current 
mediates a transition between two spinless states.
Furthermore, in the heavy mass limit the spin of the baryons
is the spin of the heavy quark,
which in the infinite mass limit remains unchanged; consequently, 
the relevant matrix elements in the region $v \sim v'$ 
can be written in terms of a form factor $B(w)$ as 
\begin{eqnarray}
 \hadm{\Xi_H  (v,s)}{\bar{s}  \gamma_\mu  u}{\Lambda_H (v',s')}%
         &=& \bar u_\Xi (v,s) u_\Lambda (v',s') B(w) (v+v')_\mu + \dots\ , \label{lambdalike}\\
 \hadm{\Xi_H  (v,s)}{\bar{s}  \gamma_\mu  \gamma_5 u}{ \Lambda_H (v',s')}%
         &=&  0 + \dots \ ,
\end{eqnarray} 
where the ellipses denote subleading contributions in the limit $v \to v'$. 

The light degrees of freedom $s_\ell$ in the  ``$\Lambda$-like''
heavy baryons form a colour anti-triplet as well as an 
anti-triplet with respect to the flavour symmetry  $SU(3)_{L+R}$ of the light quarks.
By the same argument as for the mesonic case,
one obtains a normalization statement for the form factor $B(w)$,
\begin{equation}
B(1) = 1 \ .
\end{equation} 

With the same reasoning we can obtain some insight into the form factors
for the transition from a ``$\Lambda$-like'' heavy baryon to a ``$\Sigma$-like'' heavy baryon,
i.e.\  baryons in a $h(qq')_1$ configuration. 
In the heavy mass limit, the heavy quark spin remains unchanged,
and the amplitude is determined 
by the transition of the $0^+$ state of the light degrees of freedom into a $1^+$ state.
In this way we get for $v \sim v'$, 
\begin{eqnarray} 
 \hadm{\Xi_H  (v,s)}{\bar{s}  \gamma_\mu  \gamma_5 u}{\Sigma_H (v',s')}%
     &=& \bar u_i (v,s) u_f (v',s')\epsilon_\mu A(w) + \dots \ , \label{axv-XiSigmaMT} \\ 
 \hadm{\Xi_H  (v,s)}{\bar{s}  \gamma_\mu  u}{\Sigma_H (v',s')}%
     &=&  0 + \dots  \ ,
\end{eqnarray} 
where $u_i$ ($u_f$) is the spinor of the heavy quark in the initial (final) state,
$A(w)$ is an unknown form factor,
and the ellipse again denote subleading terms. 

We have not yet specified the spin of the ``$\Sigma$-like'' heavy baryon
which can be either $1/2$ or $3/2$. 
Projecting out the relevant components 
by combining the polarization vector of the light degrees of freedom $\epsilon_\nu'$
with the heavy quark spin \cite{Mannel:1990vg, Falk:1991nq},
\begin{eqnarray} \label{spin32}
 \psi^{(3/2)}_\mu  &=& %
    \epsilon_\nu'\left[\delta_\mu^\nu  - \frac{1}{3} (\gamma_\mu + v_\mu^\prime) \gamma^\nu \right] u_f (v',s')%
       = R_\mu^{\Sigma, 3/2} (v',s') \ ,  \\
 \psi^{(1/2)}_\mu  &=&%
          \epsilon_\nu' \left[ \frac{1}{3} (\gamma_\mu + v_\mu^\prime) \gamma^\nu\right] u_f (v',s') %
       =  \frac{1}{\sqrt{3}} (\gamma_\mu + v_\mu^\prime) \gamma_5 u^{\Sigma, 1/2}  (v',s') \ ,\label{spin12}
\end{eqnarray} 
we get for the relevant matrix elements from Eq.~\eqref{axv-XiSigmaMT},
\begin{eqnarray} 
 \hadm{\Xi_H  (v,s)}{\bar{s}  \gamma_\mu  \gamma_5 u}{\Sigma_H^{(3/2)} (v',s')} &=& %
        \bar{u}_\Xi (v,s) R_\mu^{\Sigma, 3/2} (v',s')   A(w) + \dots \ , \label{sigmalike32}  \\ 
 \hadm{\Xi_H  (v,s)}{\bar{s}  \gamma_\mu  \gamma_5 u}{\Sigma_H^{(1/2)} (v',s')} &=& 
         \frac{1}{\sqrt{3}} \bar{u}_\Xi (v,s) (\gamma_\mu + v_\mu^\prime) \gamma_5 u^{\Sigma, 1/2}  (v',s') A(w) + \dots \label{sigmalike12}  \ ,
\end{eqnarray}  
where $R_\mu^{\Sigma, 3/2} $ is the Rarita-Schwinger field for the spin $3/2$ baryon
and $u^{\Sigma, 1/2} $ is the spinor for the  spin 1/2 baryon.
Note also that we have replaced the heavy quark spin of the initial state 
with the one of the ``$\Lambda$-like'' heavy baryon of the initial state. 

Close to $w = 1$ we can replace $A(w)$ by $A(1)$ however,
in this case we do not have a normalization statement, 
since the axial current generates a broken symmetry.
However, due to the heavy quark's spin symmetry we get the same factor 
$A(1)$ for both the spin $1/2$ and the spin $3/2$ case. 

Finally, the heavy $\Omega$-baryons also decay weakly,
so we also have the case of a colour-antitriplet $1^+$
state decaying into a heavy $\Xi$ or $\Xi^\prime$ baryon.
For the case of a $1^+ \to 0^+$ transition we get the 
same structure as for the $0^+ \to 1^+$ (up to complex conjugation),
while the case $1^+ \to 1^+$ needs a new discussion. 

For this we start again form the heavy mass limit and note that the heavy quark spin remind unchanged.
The underlying $1^+ \to 1^+$ transition via a vector current is usually described
in terms of six form factors out of which five vanish as $v \to v'$.
The transition amplitudes via the axial vector has to have a Levi-Civita-tensor and 
hence will vanish for $v \to v'$.
To this end we get in terms of a form factor $C(w)$ 
\begin{eqnarray} 
 \hadm{\Omega_H  (v,s)}{\bar{s}  \gamma_\mu  u}{\Xi_H^{(\prime,*)} (v',s')} &=& %
     \bar{u}_i (v,s) u_f (v',s') (\epsilon^\ast \cdot \epsilon^\prime) (v_\mu + v_\mu^\prime)  C(w) + \dots \ , \\ 
 \hadm{\Omega_H  (v,s)}{\bar{s}  \gamma_\mu  \gamma_5 u}{\Xi_H^{(\prime,*)} (v',s')} &=& %
     0 + \dots \ . 
\end{eqnarray} 
Again we have not yet specified the total spin of the baryons.
While the initial $\Omega_H$ will have total spin $1/2$,
the final states can either be spin $1/2$ or $3/2$.
Using Eqs.~\eqref{spin32} and \eqref{spin12},
we can project out the relevant components
and obtain    
\begin{eqnarray}
 &&\hadm{\Omega_H(v,s)}{\bar s\gamma_\mu u}{\Xi_H^{(3/2)}(v',s')} \notag\\
 &&=   - \frac{1}{\sqrt3}\bar u_\Omega (v,s)   \gamma_5 \left(\gamma^\alpha+v^\alpha\right) %
           R^{\Xi, 3/2}_\alpha (v',s') \left(v+v'\right)_\mu C(w)+\dots \ , \label{xilike32}\\
 &&\hadm{\Omega_H(v,s)}{\bar s\gamma_\mu u}{\Xi_H^{(1/2)}(v',s')}\notag \\ 
 &&=    -  \frac{1}{3}\bar u_\Omega (v,s)  \gamma_5 \left(\gamma^\alpha+v^\alpha\right) 
             \left(\gamma_\alpha +v_\alpha'\right)\gamma_5 u_\Xi(v',s') \left(v+v'\right)_\mu C(w)+\dots\label{xilike12} \ .
\end{eqnarray}
With the same arguments as above, we can replace $C(w)$ by $C(1)$
in the limit $w\to 1$.  Since the transition proceeds through the vector current, and the light quark states in the initial and 
final state belong to the same $SU(3)_{L+R}$ multiplet, we infer 
\begin{equation}
C(1) = 1\ .
\end{equation} 
\subsection{Semi-electronic decays with conserved heavy flavour} 
\noindent
In this section we will calculate the decay rates of heavy-flavour conserving semi-leptonic decays. 
Table~\ref{HFCons} lists all the possible semi-electronic decays of bottom and charm hadrons.   The differential decay rates
for exclusive semileptonic decays are in general given by 
\begin{equation}
\frac{d \Gamma}{d w} = \frac{G_F^2 M^5}{192 \pi^3} |V_{\rm CKM}|^2 \sqrt{w^2-1} \ P(w)    \ ,
\end{equation} 
where
\begin{equation}
P(w) =  H_{\mu \nu} (v,v') L^{\mu \nu} (v,v')  \ ,
\end{equation} 
with $H_{\mu\nu}$ and $L_{\mu\nu}$ are
the hadronic and leptonic tensors, respectively.

The integration over $w$ can be performed when setting $w=1$ in the hadronic form factors. To this end, 
it is useful to expand in the small velocity difference 
\begin{equation*}
v' = v - \Delta \, , \quad \Delta= v-v' \ .
\end{equation*}
The leptonic tensor becomes for $q=Mv-mv' = (M-m)v + m \Delta$, neglecting the electron mass
\begin{eqnarray} \label{ExpLept} 
   L_{\mu \nu} &=& g_{\mu \nu} q^2 - q_\mu q_\nu \\  
               &=& (M-m)^2 (g_{\mu \nu}  - v_\mu v_\nu)%
                   -  2 M m \, g_{\mu \nu} (w-1) \notag \\ 
               &&  - m (M-m) (\Delta_\mu v_\nu + v_\mu \Delta_\nu) %
                   - m^2 \Delta_\mu \Delta_\nu\ . \notag
\end{eqnarray} 

We shall compute the total rate,
including only the leading term in the mass difference $(M-m)$.
The integration over $w$ yields the expressions
\begin{eqnarray} \label{I0}
\int\limits_1^{w_{\rm max}} dw \ \sqrt{w^2-1}  &=& \frac{(M-m)^3}{3 M^3} 
   + {\cal O} \left(\frac{(M-m)^4}{M^4} \right) \ , \\
\int\limits_1^{w_{\rm max}} dw \ (w-1) \sqrt{w^2-1}  &=&  \frac{(M-m)^5}{10 M^5} 
+ {\cal O} \left(\frac{(M-m)^6}{M^6} \right) \ , \label{I1} \\
\int\limits_1^{w_{\rm max}} dw \ (w-1)^2 \sqrt{w^2-1}  &=&  \frac{(M-m)^7}{28 M^7} 
+ {\cal O} \left(\frac{(M-m)^8}{M^8} \right) \ ,  \\  
\int\limits_1^{w_{\rm max}} dw \ (w-1)^n \sqrt{w^2-1}  &=&  {\mathcal O} \left( \frac{(M-m)^{2n+3}}{M^{2n+3}} \right) \ , 
\end{eqnarray} 
which show that a one power of $(w-1)$ in the differential rate
counts as two powers of $(M-m)$ in the total rate.
Hence, looking at the expansion \eqref{ExpLept} of the leptonic tensor
we note that the leading terms of $L_{\mu \nu}$ 
are already of order $(M-m)^2$.
Note that, depending on the 
hadronic tensor, even the last term involving $\Delta_\mu \Delta_\nu$ needs to be kept,
since $\Delta^2 = 2 v \cdot \Delta = 2 (1-w) \sim (M-m)^2$. 

For the hadronic tensor this means that we need to include only the leading term with $\Delta = 0$,
which is in all cases of order $(M-m)^0$. 
The simplest process is the decay $0^- \to 0^-$ between ground states,
where we have a light quark transition in the background field of the heavy quark.
Using the discussion from the previous section, 
we insert for the hadronic tensor Eq.~\eqref{mesonFF} 
\begin{equation}
H_{\mu \nu}  =  4 M^2 v_\mu v_\nu  \ .
\end{equation} 
Inserting the integral \eqref{I1}, we get 
\begin{equation} \label{dG0m0m}
\Gamma^{0^-\to0^-} =  \frac{G_F^2}{60 \pi^3} |V_{\rm CKM}|^2 (M - m)^5\ .
\end{equation} 

For the transition $0^- \to 1^-$  mesons we obtain for the hadronic tensor from \eqref{mesonFFs}
\begin{eqnarray}
  H_{\mu \nu}  &=&  4 M^2  |\Phi_A (1)|^2 \sum_{\rm Pol} \epsilon_\mu \epsilon_\nu  \notag\\
               &=&  4 M^2  |\Phi_A (1)|^2 (g_{\mu \nu}  - v_\mu v_\nu) \ . 
\end{eqnarray} 
Using the integrals \eqref{I0} and \eqref{I1},
and keeping only the leading order,
we get using \eqref{mesonFFs} for the total decay rate
\begin{equation}\label{dG0m1m}
\Gamma^{0^-\to1^-} =  \frac{G_F^2}{20 \pi^3} |V_{\rm CKM}|^2 (M - m)^5   |\Phi_A (1)|^2\ . 
\end{equation} 

For our numerical estimates shown in Table~\ref{mesondec} 
we shall set $ \Phi_A (1) = 1$. Note that the result for  $ \Phi_A (1) = 1$
just reflects spin counting, 
furthermore, the sum of the two rates is just the total, spin-summed decay rate 
for the spin 1/2 light system decaying in the colour background of the heavy quark. 

Table~\ref{mesondec} lists the rates and
the branching ratios $(\mathcal B)$ for the mesonic semileptonic decays. 
Note that the $D^+ \to D^0$ decay is a $d \to u$ transitions, 
while all other decays are $s \to u$.
\begin{table}[t!]
\caption{\label{mesondec} Branching ratios for semileptonic meson decays as discussed in the text.}
\vspace*{4mm} 
\begin{tabular*}{\textwidth}{l@{\extracolsep{\fill}}ll}\hline\hline
  Mode                       & Decay Rate [GeV]      & Branching Ratio  \\\hline 
  $D^+   \to D^0 e^+ \nu$    & $1.72\times10^{-25}$  & $2.71\times10^{-13}$ \\
  $D_s^+ \to D^0 e^+ \nu$    & $4.40\times10^{-20}$  & $3.34\times10^{-8}$  \\
  $B_s^0 \to B^- e^+\nu$     & $1.90\times10^{-20}$  & $4.37\times10^{-8}$  \\
  $B_s^0 \to B^{*-} e^+\nu$  & $1.38\times10^{-21}$  & $3.17\times10^{-9}$  \\\hline \hline                                         
 \end{tabular*}
 \end{table}

With the same method we can discuss the semi-electronic decays of heavy baryons.
As discussed above the light degrees of freedom are more complicated in this case.
For this reason we introduce the notation, where the superscript
denotes the spin-parity of the baryon transitions,
while the subscripts denote the spin-parity of the corresponding transition of the 
light degrees of freedom.

For the decays  of 
the type $\Xi \to \Lambda e \bar{\nu}$ where the light degrees of freedom are in a spineless state,
we obtain using \eqref{lambdalike} 
\begin{equation}
\Gamma^{1/2^+\to1/2^+}_{0^+\to0^+} = \frac{G_F^2 |V_{\rm CKM}|^2 }{60 \pi^3} (M-m)^5\ ,      
\end{equation} 
where we have used the form factors obtained in the previous subsection.  Note that this is the same 
result as for the mesonic $0^- \to 0^-$ transition, which is not surprising, since this is just a spinless 
system of light degrees of freedom decaying in the colour-background of the heavy quark. 

For the final states with a ``$\Sigma$-like" baryon we obtain
from Eqs.~\eqref{sigmalike12} and \eqref{sigmalike32},
\begin{eqnarray}
 \Gamma^{1/2^+\to3/2^+}_{0^+\to1^+}&=&%
   \frac{G_F^2\left|V_{\rm CKM}\right|^2}{30\pi^3} (M-m)^5 |A(1)|^2 \ , \\
 \Gamma^{1/2^+\to1/2^+}_{0^+\to1^+} &=&%
   \frac12\Gamma^{1/2^+\to3/2^+}_{0^+\to1^+} \ ,   
\end{eqnarray}
where we again note that the sum of the two rates is just the result we obtained for the mesonic $0^- \to 1^-$ 
transition. Again this is due to spin counting, since in both decays we observe a transition of a light $0^+$ state into 
a light $1^+$ state, however, with different spin combinations with the heavy quark.

Finally, utilizing ~\eqref{xilike32} and \eqref{xilike12}
we find for final states with a ``$\Xi$-like" baryon,
\begin{eqnarray}
 \Gamma^{1/2^+\to1/2^+}_{1^+\to1^+} &=&%
    \frac{G_F^2\left|V_{\rm CKM}\right|^2}{15\pi^3} (M-m)^5 \ ,  \\ 
 \Gamma^{1/2^+\to3/2^+}_{1^+\to1^+}&=& {\mathcal O}( [M-m]^7 ) \ . %
\end{eqnarray}
where the last line means that this transition
has an additional suppression factor $(M-m)^2 / M^2$ compared to the other decays, 
the rates of which are all of the order $G_F^2 (M-m)^5$.
Since we only considered the leading terms of the form factors for $v \sim v'$, 
we cannot obtain a result for these decay on the basis of the discussion in section~\ref{ssec:FF}. 
 
For our numerical estimates we shall set $|A(1)|^2 = 1$;
in Table~\ref{baryondec} we list the branching ratios
for possible semi-electronic baryon decays with conserved heavy flavour.  

\begin{table}[t!]
\caption{\label{baryondec} Decay rates and branching ratios for semi-electronic baryon decays as explained in the text.}
\vspace*{4mm} 
\begin{tabular*}{\textwidth}{l@{\extracolsep{\fill}}ll}\hline\hline
  Mode                       & Decay Rate [GeV]%
                             & Branching Ratio  \\\hline 
  $\Xi_c^0  \to \Lambda_c^{+}  e^- \bar\nu$
                             & $7.91\times10^{-19}$
                             & $1.35\times10^{-7}$  \\
  $\Xi_c^0  \to \Sigma_c^{+}  e^- \bar\nu$ 
                             & $6.97\times10^{-24}$
                             & $1.19\times10^{-12}$ \\
  $\Xi_c^+\to \Sigma_c^{++} e^- \bar\nu$ 
                             & $3.74\times10^{-24}$
                             & $1.26\times10^{-12}$ \\
  $\Omega_c^0  \to \Xi_c^{+}  e^- \bar{\nu}$
                             & $2.26\times10^{-18}$
                             & $2.36\times10^{-7}$  \\
  $\Omega_c^0  \to \Xi_c^{\prime +}  e^- \bar{\nu}$%
                             & $3.63\times10^{-19}$
                             & $3.81\times10^{-8}$  \\
  $\Omega_c^0  \to \Xi_c^{* +}  e^- \bar{\nu}$%
                             & $1.49\times10^{-29}$
                             & $1.57\times10^{-18}$  \\\hline                       
  $\Xi_b^-  \to \Lambda_b^{0}  e^- \bar{\nu} $
                             & $6.16\times10^{-19}$
                             & $1.46\times10^{-6}$  \\ 
  $\Omega_b^-  \to \Xi_b^0  e^- \bar{\nu} $%
                             & $4.05\times10^{-18}$
                             & $6.78\times10^{-6}$  \\
  $\Omega_b^-  \to \Xi_b^{* 0}   e^- \bar{\nu} $%
                             & $3.27\times10^{-28}$
                             & $5.47\times10^{-16}$  \\\hline\hline                                          
 \end{tabular*}
\end{table}

\subsection{Semi-Muonic Decays}
\noindent
For a few of the baryonic decays phase space is large enough to allow for semi-muonic decay.
In this case we have to take into account the mass $m_\mu$ of the muon in the leptonic tensor 
\begin{equation} \label{LTmass}
L_{\mu \nu} = \frac{(q^2-m_\mu^2)^2(2 q^2 + m_\mu^2)}{2 q^2}  g_{\mu \nu} 
- \frac{(q^2)^3-3 m_\mu^4 q^2 +2 m_\mu^6}{(q^2)^3} q_\mu q_\nu   \ , 
\end{equation}   
with $q=Mv-mv'$. 

The muon mass is of the same order as the mass difference $(M-m)$ between the initial and the final state 
baryon, and thus an expansion in $(M-m)$ as in the massless case is spoiled by the presence of the ratio 
$m_\mu / (M-m) \sim {\cal O}(1)$. 
Hence we perform the integration over the phase space after contracting the leptonic tensor \eqref{LTmass}  
with the hadronic tensors taken at $v=v'$ without the expansions \eqref{I0} and \eqref{I1} performed in the massless case. 
The results for the rates and branching fractions are shown in  Table~\ref{tab:SemiMu}. 
It is interesting to note that the branching ratios for the semi-muonic channels are not that much smaller as it is 
suggested by phase space; this effect is due to the presence of the muon mass in the leptonic tensor. 

\begin{table}[b!]
\caption{\label{tab:SemiMu} Decay rates and branching ratios for semi-muonic baryon decays as explained in the text.}
\vspace*{4mm} 
\begin{tabular*}{\textwidth}{l@{\extracolsep{\fill}}ll}\hline\hline
  Mode                       & Decay Rate [GeV]%
                             & Branching Ratio  \\\hline 
  $\Xi_c^0  \to \Lambda_c^{+}  \mu^- \bar\nu$
                             & $ 1.3 \times 10^{-19} $
                             & $ 2.3 \times 10^{-8}  $  \\
  $\Omega_c^0  \to \Xi_c^{+}  \mu^- \bar{\nu}$
                             & $ 7.1 \times 10^{-19} $
                             & $ 7.4 \times 10^{-8} $   \\
   $\Omega_c^0  \to \Xi_c^{+ \prime}  \mu^- \bar{\nu}$
                             & $ 1.0 \times 10^{-21} $
                             & $ 1.1 \times 10^{-10} $                            
                             \\ \hline                    
  $\Xi_b^-  \to \Lambda_b^{0}  \mu^- \bar{\nu} $
                             & $9.1 \times 10^{-20} $
                             & $2.2 \times 10^{-7}$  \\ 
  $\Omega_b^-  \to \Xi_b^0  \mu^- \bar{\nu} $%
                             & $1.7 \times 10^{-18}$
                             & $2.8 \times 10^{-6}$  \\ \hline\hline                                          
 \end{tabular*}
\end{table}

\subsection{Non-leptonic (pionic) decays} 
The non-leptonic decays with conserved heavy flavour are an interesting QCD laboratory for light quarks and gluons 
moving in the background field of the heavy quark; for this reason they have been studied already to some extend
in \cite{Voloshin:2000et,Li:2014ada} and we mainly update these analyses. 

The relevant effective Hamiltonian is the usual $\Delta S = \pm 1$
weak-transition Hamiltonian,
which reads 
\begin{eqnarray}
 \ham_{\rm eff}^{(l)} &=& \frac{4 G_F}{\sqrt{2}} V_{us} V_{ud}^\ast \sum_i C_i O_i  \\ \notag
                            &=& \frac{4 G_F}{\sqrt{2}} V_{us} V_{ud}^\ast
                                      \left[ C_+ (\bar{s}_L \gamma_\mu u_L)(\bar{u}_L \gamma^\mu d_L)  
                                           + C_- (\bar{s}_L \gamma_\mu d_L)(\bar{u}_L \gamma^\mu u_L)  \vphantom{\int} \right]
                                           + \cdots \ ,
\end{eqnarray} 
where $C_+ \approx 1.3$ and $C_- \approx -0.6$,
and where we omitted all contributions with 
very small Wilson coefficients $C_i$. 

This part of the effective Hamiltonian is sufficient for the heavy-flavour conserving decays of bottom baryons; 
however, as has been pointed out by Voloshin \cite{Voloshin:2000et} there is another relevant contribution 
for charmed baryons  
\begin{equation}
\ham_{\rm eff}^{(c)}  = \frac{4 G_F}{\sqrt{2}} V_{cs} V_{cd}^\ast%
                               \left[ C_+ (\bar{s}_L \gamma_\mu c_L)(\bar{c}_L \gamma^\mu d_L) %
                                    + C_- (\bar{s}_L \gamma_\mu d_L)(\bar{c}_L \gamma^\mu c_L)  \vphantom{\int} \right]
\end{equation} 
generating a difference in the decay amplitudes for the heavy-flavour conserving decays of charm baryons compared to the 
corresponding amplitudes for bottom baryons.

Since the phase space of the pion is rather small, one may use the soft pion limit to gain some further 
insight \cite{Voloshin:2000et}. The soft pion theorem allows us to write
\begin{equation} \label{SPT}
\left\langle B_i \left|  \ham_{\rm eff} \right| B_f \pi^a (\vec{p}_\pi = 0)  \right\rangle = \frac{\sqrt{2}}{f_\pi} 
\left\langle B_i \left| \left[ \ham_{\rm eff} \, , \, Q_5^a \right] \right| B_f  \right\rangle \ , 
\end{equation}  
where $Q^a_5$ is the axial charge corresponding to the pion 
\begin{eqnarray}
Q_5^+ &=& \int d^3 \vec{x} \, \bar{u}(x) \gamma_0 \gamma_5 d(x)\ ,  \quad Q_5^- = (Q_5^+)^\dagger\ , \\ 
Q_5^0 &=& \frac{1}{\sqrt{2}} \int d^3 \vec{x} \  
            \left( \bar{u}(x) \gamma_0 \gamma_5 u(x) - \bar{d}(x) \gamma_0 \gamma_5 d(x) \right)\ ,
\end{eqnarray} 
and $f_\pi \sim 130$~MeV is the pion decay constant. 

It has been shown in Ref.~\cite{Voloshin:2000et}
that in this limit the transitions are dominated by the $S$ wave 
and are purely $\Delta I  =1/2$. Thus to a very good approximation one has   
\begin{eqnarray}
 \hadm{\Xi_c^+}{\ham_{\rm eff}}{\Lambda_c^+ \pi^0} &=&%
    \frac{1}{\sqrt{2}}
       \hadm{\Xi_c^0}{\ham_{\rm eff}}{\Lambda_c^+ \pi^-}\ ,  \\
 \hadm{\Xi_b^0}{\ham_{\rm eff}}{\Lambda_b^0 \pi^0}&=& %
    \frac{1}{\sqrt{2}}
       \hadm{\Xi_b^-}{\ham_{\rm eff}}{\Lambda_b^0 \pi^-} \ ,
\end{eqnarray} 
and 
\begin{eqnarray}
 \hadm{\Omega_c^0}{\ham_{\rm eff}}{\Xi_c^0\pi^0} &=& %
    \frac{1}{\sqrt{2}}
        \hadm{\Omega_c^0}{\mathcal \ham_{\rm eff}}{\Xi_c^+\pi^-}\ , \\ 
 \hadm{\Omega_b^-}{\ham_{\rm eff}}{\Xi_c^-\pi^0} &=& %
     \frac{1}{\sqrt{2}} 
        \hadm{\Omega_b^0}{\mathcal \ham_{\rm eff}}{\Xi_b^0\pi^-} \ .
 \end{eqnarray}  

We first consider the decays of the bottom baryons for which we do not need to take into account $ \ham_{\rm eff}^{(c)} $. 
Clearly the matrix element of the weak Hamiltonian is difficult to estimate,
and we will be able to make only rather qualitative statements.
We shall approach this problem from the point of view of the head quark limit: 
the heavy quark completely decouples from the process,
leaving a weak decay of a di-quark system in the background field of the heavy quark. 
To this end, the amplitude for the $\Xi_b^- \to \Lambda_b^0 \pi^-$ transitions may be written as 
\begin{eqnarray}  \label{Adef} 
 \hadm{\Xi_b^-}{\ham_{\rm eff}}{\Lambda_b^0 \pi^-}  &=&%
          \bar{u}_\Xi (v) u_\Lambda (v') \hadm{(sd)_0}{\ham_{\rm eff}}{ (ud)_0  \pi^-}_{\rm ext} \notag  \\  
      &\equiv& \bar{u}_\Xi (v) u_\Lambda (v')  {\mathcal A} ((sd)_0 \to (ud)_0  \pi^-) \ ,
\end{eqnarray} 
where the subscript ``ext'' means that the transition is taking place in the color-background field of the heavy quark, 
and $(qq')_{s_\ell}$ denotes a di-quark with total spin $s_\ell$.
Using this notation, the decay rate for $\Xi_b^- \to \Lambda_b^0 \pi^-$ becomes
\begin{equation} \label{Xirate}
   \Gamma (\Xi_b^- \to \Lambda_b^0 \pi^-) = %
      \frac{\sqrt{[M^2-(m-m_\pi)^2][M^2-(m+m_\pi)^2]}}{16  \pi M^3}  
      \ \left| {\mathcal A} ((sd)_0 \to (ud)_0  \pi^-)  \right|^2  \ .         
\end{equation}

Unfortunately nothing is known about the matrix element for the di-quark decay, so we only can set this into 
a relation with with the typical amplitudes for a weak transition. If we consider the weak decays of a pseudo scalar 
meson into a final state of two pseudo scalar mesons $M \to M_1 + M_2$, we get 
\begin{equation}
 \hadm{ M }{\ham_{\rm eff}}{M_1 \, M_2} = 2M\ V_{\rm CKM} \ a_{\rm weak} \ ,  
\end{equation} 
where $M$ is the mass of the initial state and the usual relativistic normalization is used.
The value for $| a_{\rm weak}|$ covers only a limited
range when scanning over the weak decays 
of $B$, $D$ and $K$ mesons
\begin{equation} 
|a_{\rm weak}|  \sim (1 \cdots 2) \times 10^{-6} \ .
\end{equation} 
If the amplitude for the di-quark transition is of the same order of magnitude, we estimate 
\begin{equation}
{\mathcal A} ((sd)_0 \to (ud)_0  \pi^-)  \sim  2 M \  V_{us}V_{ud}^\ast\ a_{\rm weak} \ ,
\end{equation} 
where the spinors in \eqref{Adef} are assumed to be normalized non-relativistically. 
Inserting the numbers, 
we obtain the estimates given in Table~\ref{pionbaryondec}. 
 
In the same spirit we can deal with the decay $\Omega_b^- \to \Xi_b^0 \pi^-$.
However, here the di-quark of the initial state has spin one,
so we get 
\begin{equation} 
 \hadm{\Omega_b^-}{\ham_{\rm eff}}{\Xi_b^0 \pi^-}  = %
       \frac{1}{\sqrt{3}} \sum_\lambda \bar{u}_\Omega (v) \gamma_5 \fmslash{\epsilon} (\lambda) \, u_\Xi (v') 
       \hadm{(ss)_1, \lambda}{\ham_{\rm eff}}{ (us)_0  \pi^-}_{\rm ext} \ ,
\end{equation} 
where $\lambda = \pm , 0 $ are the polarizations of the vector di-quark in the initial state.
The matrix element will have the form 
\begin{equation}
 \hadm{(ss)_1, \lambda}{\ham_{\rm eff}}{ (us)_0  \pi^-}_{\rm ext} =%
  (\epsilon^\ast(\lambda) \cdot v')  {\mathcal A}\left ( (ss)_1 \to (us)_0  \pi^-\right) \ ,
\end{equation} 
so we obtain, assuming equal amplitudes of all helicities
\begin{equation} 
 \hadm{\Omega_b^-}{\ham_{\rm eff}}{\Xi_b^0 \pi^-}  = 
 \frac{1}{\sqrt{3}} (1+vv')  \bar{u}_\Omega (v) \gamma_5  u_\Xi (v')  \, 
       {\mathcal A} \left( (ss)_1 \to (us)_0  \pi^-\right) \ .
\end{equation} 
From this we obtain for the rate
\begin{equation}
\Gamma (\Omega_b^- \to \Xi_b^0 \pi^-) = %
   \frac{  ([M^2-(m-m_\pi)^2][M^2-(m+m_\pi)^2])^{3/2}}{192  \pi M^7}  
       \ | {\mathcal A} \left((ss)_1 \to (us)_0  \pi^-\right)  |^2   \ .        
\end{equation} 
Assuming agin that the corresponding amplitude is of the order 
\begin{equation}
 {\mathcal A} \left( (ss)_1 \to (us)_0  \pi^-\right)  \sim %
  2 M \  V_{us}V_{ud}^\ast \ a_{\rm weak} \ , 
\end{equation} 
we obtain the numbers of in Table~\ref{pionbaryondec}. 

The pionic heavy flavour conserving of charmed hadrons involve
also the part  $\ham_{\rm eff}^{(c)}$ of the effective weak Hamiltonian.
We first consider the decay $\Xi_c^0 \to \Lambda_c^+ \pi^-$.
Making use of the soft-pion theorem \eqref{SPT}, we obtain
\begin{equation} 
 \hadm{\Xi_c^0  }{\ham_{\rm eff}^{(c)} }{\Lambda_c^+ \pi^-} =%
 - \frac{\sqrt{2}}{f_\pi}  
    \hadm{\Xi_c^0  }{\left[ \ham_{\rm eff}^{(c)} \ , \ Q_5^- \right] }{\Lambda_c^+  } \ .
\end{equation} 
In the infinite mass limit, only the vector current contributes, so we get
\begin{equation}  \label{Hc}
 \hadm{\Xi_c^0  }{\ham_{\rm eff}^{(c)} }{\Lambda_c^+  \pi^-} =%
   - \frac{G_F}{f_\pi} V_{cs} V_{cd}^\ast 
      \hadm{\Xi_c^0  }{(\bar{c} \gamma_\mu c)(\bar{s} \gamma^\mu d)  }{\Lambda_c^+ } \ .
\end{equation} 
It has been shown in \cite{Voloshin:2000et} that one may obtain information on these matrix elements from the 
lifetime differences of charmed baryons, assuming light-quark flavour symmetry.
The number found in \cite{Voloshin:2000et} is 
\begin{equation}
 \hadm{\Xi_c^0  }{\ham_{\rm eff}^{(c)} }{\Lambda_c^+  \pi^-} \sim - 2 M \times \left(  5.4 \times 10^{-7}  \right)\ ,
\end{equation} 
with an uncertainty of about 50\%. 

Here we shall try to understand the the anatomy of these matrix elements
a bit further in order to also include an estimate for the pionic decays of the $\Omega_c$.
To this end we note that in the infinite mass limit we have to match \eqref{Hc}
on static heavy quarks moving with the same velocity $v$,
since in the soft-pion limit the heavy quark velocity does not change.
As a consequence we get 
\begin{equation}  
 \hadm{\Xi_c^0  }{\ham_{\rm eff}^{(c)} }{\Lambda_c^+  \pi^-} = %
   2 M \ \bar{u}_\Xi (v) u_\Lambda (v')  
      \left(\frac{G_F}{f_\pi} V_{cs} V_{cd}^\ast \ \mu^3 \right) \ ,
\end{equation} 
where $\mu$ is a nonperturbative hadronic scale of the order of a few hundred MeV,
which is related to the wave functions of the constituents at the origin. 
In fact, from the result of \cite{Voloshin:2000et} we infer $\mu \sim 300$ MeV.  

The full amplitude for $\Xi_c^0 \to \Lambda_c^+ \pi^-$ consists of the two contributions
from $\ham_{\rm eff}^{(l)}$ and $\ham_{\rm eff}^{(c)} $.
However, nothing is known about the relative phase $\phi$ of these two contributions,
so one gets
\begin{equation}  
 \hadm{\Xi_c^0  }{H_{\rm eff} }{\Lambda_c^+  \pi^-} = 2 M \, \bar{u}_\Xi (v) u_\Lambda (v)  
 \left( V_{us}V_{ud}^\ast\ e^{i \phi}  a_{\rm weak}  +  \frac{G_F}{f_\pi} V_{cs} V_{cd}^\ast \, \mu^3 \right)  \, . 
\end{equation} 
Assuming constructive inference we get an upper limit for the decay rates and branching fractions for these decays,
which is shown in Table~\ref{pionbaryondec}.

When considering the decays $\Omega_c \to \Xi_c \pi$
we need to take into account that the the light degrees of freedom in the $\Omega_c$ 
are in a spin-1 state.
To this end, the relevant matrix element takes the form 
\begin{equation}  
 \hadm{\Omega_c^0  }{\ham_{\rm eff}^{(c)} }{\Xi_c^+  \pi^-} =%
   2 M \frac{1}{\sqrt{3}} \ \bar{u}_\Omega (v) \gamma_5 (\gamma_\alpha + v_\alpha) u_\Lambda (v')\ W^\alpha \ ,
\end{equation} 
where $W^\alpha$ describes the decay of the vector di-quark into a scalar di-quark under the emission of a pion.
Note that we have not yet set $v = v'$, 
since the vector-spinor object for the $\Omega_c$ is transverse
and the amplitude would vanish for $v = v'$.
In our estimates, we assume for the quantity $W^\alpha$   
\begin{equation}
W_\alpha = v^\prime_\alpha  \left(\frac{G_F}{f_\pi} V_{cs} V_{cd}^\ast \  \mu^3 \right) \ ,
\end{equation} 
with the same hadronic parameter $\mu$.
This yields
\begin{equation}
  \hadm{\Omega_c^0  }{\ham_{\rm eff}^{(c)} }{\Xi_c^+  \pi^-} =%
     \frac{2M}{\sqrt{3}} (1+vv') \  \bar{u}_\Omega (v) \gamma_5 u_\Lambda (v')%
       \left(\frac{G_F}{f_\pi} V_{cs} V_{cd}^\ast \  \mu^3 \right) \ . 
\end{equation} 

Inserting numbers we can estimate the contribution from $\ham_{\rm eff}^{(c)}$.
As above, the relative phases of the  contributions from 
$\ham_{\rm eff}^{(c)}$ and $\ham_{\rm eff}^{(l)}$ are to known,
we end up with a large uncertainty in our prediction.

\begin{table}[t!]
\caption{\label{pionbaryondec} Branching ratios for pionic baryon decays as explained in the text.}
\vspace*{4mm} 
\begin{tabular*}{\textwidth}{l@{\extracolsep{\fill}}rr}\hline\hline
  Mode                       & Decay Rate [GeV]
                             & Branching Ratio \\\hline 
  $\Xi_b^-  \to \Lambda_b^{0}  \pi^-$
                             & $(0.8 \cdots 3.2)\times 10^{-15}$
                             & $(1.9 \cdots 7.6)\times 10^{-3}$ \\   
  $\Xi_b^0  \to \Lambda_b^{0}  \pi^0$
                             & $(0.4  \cdots 1.7)\times 10^{-15}$
                             & $(0.9  \cdots 3.7)\times 10^{-3}  $   \\   
  $\Omega_b^-  \to \Xi_b^0  \pi^-$
                             & $(0.7  \cdots 2.6)\times 10^{-18} $
                             & $(1.1  \cdots 4.3)\times 10^{-6} $  \\        
  $\Omega_b^-  \to \Xi_b^-  \pi^0$
                             & $(0.3  \cdots 1.3) \times10^{-18}$
                             & $(0.6  \cdots 2.2) \times 10^{-6}$  \\\hline
  $\Xi_c^0  \to \Lambda_c^{+}  \pi^-$
                             & $< 1.7 \times 10^{-14}$
                             & $<3 \times 10^{-3}$   \\
  $\Xi_c^+  \to \Lambda_c^{+}  \pi^0$
                             & $< 8.8 \times 10^{-15}$
                             & $< 6 \times 10^{-3}$  \\
  $\Omega_c^0  \to \Xi_c^{+}  \pi^-$ 
                             & $< 3.5 \times10^{-17}$
                             & $< 3.7 \times 10^{-6}$  \\
  $\Omega_c^0  \to \Xi_c^{0}  \pi^0$ 
                             & $<1.8 \times10^{-17}$
                             & $<1.1  \times10^{-6}$ \\\hline\hline                                           
 \end{tabular*}
\end{table}

\section{Summary} 
Heavy flavour conserving weak decays will very likely not advance our insight into weak interactions;
however, they may be an interesting QCD laboratory for the study of 
light-quark systems in the colour-background field of a heavy quark.
While for heavy mesons this mainly is the decay of a light quark in such a background field,
the situation for a heavy baryon may be more interesting in this respect,
since the light degrees of freedom form a more complicated system. 

The semi-electronic modes are under reasonable theoretical control
and thus may serve as a benchmark test for the pionic modes. 
Like in non-leptonic kaon processes, naive factorization will probably not work,
but the numbers obtained in this way may give a hint of the 
size of the branching fractions.
Here it will be interesting to see, 
if some patterns observed in the kaon system also appear,
if the light-quark systems decay in a colour background field. 

One obvious disadvantage of these decays is their suppression
through the small phase space. Relative to the major decay modes, 
these decays suffer from a suppression factor $(M-m)^5 / M^5$ for the semi-electronic modes,
and  the phase space suppression for the pionic modes is numerically about the same.
This leaves branching fractions of the oder of  $10^{-6}$ in the best cases,
typically $10^{-7}$ to  $10^{-8}$. This makes the investigation of these decays a challenge 
for the $B$ physics experiments.

\subsection*{Acknowledgements}
We thank Mike Versterinen for a discussion which renewed out interest in this type of decays.
We are particularly grateful to Misha Voloshin who directed out attention to \cite{Voloshin:2000et,Li:2014ada}
where the pionic decays had been discussed  already exhaustively. 
This work was supported by the german research foundation DFG in the frame of the research unit FOR 1873.

\end{document}